\documentclass[11pt]{article}
\usepackage{latexsym}

\def\hb{\hbox}
\def\rot{\hb{rot}}
\def\h{\hb{\bf h}}

\def\r{\hb{\bf r}}
\def\ds{\displaystyle }
\def\xr{\hb{\bf x+r}}
\def\v{\hb{\bf v}}

\begin{document}

\title{On triple correlations in isotropic electron
magnetohydrodynamic turbulence}
\author{Otto Chkhetiani,\thanks{%
ochkheti@mx.iki.rssi.ru} \\
Space Research Institute, RAS,\\
Profsoyuznaya 84/32, Moscow, 117997, Russia}
\date{}
\maketitle

\begin{abstract}
The evolution of the correlation characteristics in
three-dimensional isotropic electron magnetohydrodynamic
turbulence is investigated. Universal exact relations between the
longitudinal and  transverse two-point triple correlations of the
components of the fluctuational magnetic fields and the rates of
dissipation of the magnetic helicity and energy are obtained in
the inertial range.
\end{abstract}

PACS numbers: 52.35.Bj, 52.35.Ra

All statistical theories of turbulence take into account the
well-known exact Kolmogorov result --- the 4/5 law \cite{kolm41}
which relates the third-order spatial longitudinal correlations of
the velocity with the rate of energy dissipation. In
magnetohydrodynamics
 a relation was obtained by Chandrasekhar \cite{chandras}.
Recently a similar relation (2/15 law) was established for
hydrodynamic turbulence with helicity \cite{chkhet96,lvov97}. The
confirmations of the 4/5 law for diverse turbulent hydrodynamic
flows are well known \cite{Monin96}. Confirmations have been
obtained for the 2/15 law \cite{chkhet96,lvov97} for helicity
\cite{biferale}. It is important to note that such accurate
relations are obtained by solving dynamical equations and are a
consequence of the conservation laws. No dimensional
considerations are employed in their derivation. The
 fundamental significance of the 4/5 law in hydrodynamics
has been examined in detail  in \cite{frisch95}.

Electronic magnetohydrodynamics (EMHD) pertains to a branch of
plasma oscillations on which the Hall term predominates
\cite{kingsep87,gordeev94} and it is a limiting case of
multicomponent  MHD, where the motion of the ions can be neglected
and the motion of the electrons
 preserves quasineutrality. In contrast to the standard
MHD case, the description (with  uniform density) can be reduced
to a single nonlinear equation for the magnetic field. The
 region of applicability of EMHD are laboratory and
industrial plasma setups, the ionosphere, the solar photosphere,
and solids \cite{gordeev94,vainshtein76}. In the 1970s the term
MHD at helicon frequencies was also used \cite{vainshtein76}. Weak
turbulence of helicons (whistler) was studied in
\cite{yakovenko69,cytovich,livshitz72}. The dynamic properties of
strong three-dimensional turbulence in EMHD have been studied in
\cite{vainshtein73}. Arguments supporting the idea that only weak
turbulence is  realized in the EMHD mechanism are presented in
\cite{kingsep87}.

EMDH is described by the equation \cite{kingsep87,gordeev94}

\begin{eqnarray}
\partial_t \h+\rot\left[\frac{\hb{\bf j}}{ne}\times \h\right]
+c\>\rot\frac{\hb{\bf j}}{\sigma}=0,\\
\hb{\bf j}=\frac{c}{4\pi}\rot\h,\qquad \div \h=0.
\end{eqnarray}
For $n=const$ and $\sigma=const$ obtain
\begin{equation}
\partial_t\h=-\frac{c}
{4\pi ne}\rot\left[ \rot\h\times \h\right]
+\frac{c^{2}}{4\pi\sigma }\Delta \h, \qquad \div\h=0. \label{eqn0}
\end{equation}
In the frequency domain this corresponds to the range
\[
w_i<w<w_e.
\]
For what follows, we introduce the notation
$\ds\hb{f}=\frac{c}{4\pi ne}$ and
$\ds\nu_m=\frac{c^2}{4\pi\sigma}$. The structure of
Eq.(\ref{eqn0}) is close to the Navier-Stokes equation for an
incompressible liquid. We can verify by direct substitution that
it conserves the energy
$$
\ds\int\limits_V\frac{\h^2}{2}dV
$$
and the helicity
$$
\ds\int\limits_V\hb{\bf ah}dV
$$ of the magnetic field.

Let us consider the free evolution of uniform and isotropic
fluctuations of the magnetic field in EMHD. Writing out the
equation for the vector potential $\hb{\bf a}=\rot^{-1}\h$ and
averaging, we obtain equations of the Karman-Howarth type for the
two-point correlation functions involving the energy and helicity
of the magnetic field:
$$
\partial_t h_{ii}=
\hb{f}\varepsilon_{ijk}\frac{\partial^{2}} {\partial r_{i}\partial
r_{m}} \left( h_{j,km}-h_{jm,k}\right) +2\nu_m\Delta_{r}h_{ii}=
$$
\begin{equation}
\hb{f}\varepsilon_{ijk}\frac{\partial^{2}}{\partial r_{i}\partial
r_{m}}\left( h_{km,j}-h_{jm,k}\right) +2\nu_m\Delta_{r}h_{ii}
\label{eqn1}
\end{equation}
\begin{equation}
{\ds
\partial_t g_{ii}=2\hb{f}\frac{\partial}{\partial r_{m}} h_{im,i}
+2\nu_m\Delta_{r} g_{ii} \label{eqn2}, }
\end{equation}
where
\begin{eqnarray}
h_{ii}=\langle h_i(\hb{\bf x})h_i(\xr)\rangle, \qquad
g_{ii}=\langle a_i(\hb{\bf x}) h_i(\xr)\rangle,\\
h_{jm,k}=\langle h_j(\hb{\bf x})h_m(\hb{\hb{\bf x}})
h_k(\xr)\rangle.
\end{eqnarray}

The right-hand sides of Eqs. (\ref{eqn1}) and (\ref{eqn2}) contain
the spatial derivatives with respect to $r$ of the rank-3
two-point correlation tensor. The general form of such a tensor,
with  allowance for the gyrotropy and incompressibility of the
magnetic field, is \cite{chandras,chkhet96,lvov97}
\begin{equation}
h_{ij,k}\left(\r\right)=V\left( \varepsilon _{jkl}r _{i}r _{l}
+\varepsilon _{ikl}r _{j}r _{l}\right) +\frac{2}{r }\partial_r Tr
_{i}r _{j}r _{k}-\left( r \partial_r T+3T\right) \left( r
_{i}\delta _{jk}+r _{j}\delta _{ik}\right) +2T\delta _{ij}r
_{k}.\label{tensor}
\end{equation}
Fluctuations of the magnetic field without helicity were
considered in \cite{chandras}. In that case the tensor $h_{ij,k}$
consists of only the first two terms, which are proportional to
the scalar $V$, which is related to the energy transfer. Taking
the helicity into account introduces additional terms which are
proportional to the product of pseudoscalar quantities and odd
 combinations of the components of the radius vector.
Formally, the solenoidal tensor (\ref{tensor})  is identical to
the analogous tensor for triple correlations of the velocity field
in hydrodynamic turbulence \cite{chkhet96}. However, in contrast
to the latter it does not change under reflection  of the
coordinates, i.e., $h_{ij,k}(-\r)=h_{ij,k}(\r)$. The properties of
homogeneous turbulence
 also imply that $h_{k,ij}(r)=h_{ij,k}(-r)$ \cite{Monin96}. Both properties
are taken into account in Eq.(\ref{eqn1}).

In what follows we shall need to examine the auxiliary tensor
$\langle\delta h_i(\hb{\bf x}|\r)\delta h_j(\hb{\bf
x}|\r)\rangle$, where $\delta \h(\hb{\bf x}|\r)=\h(\xr)-\h(\hb{\bf
x})$. In homogeneous turbulence it has the form
\[
\langle\delta h_i(\hb{\bf x}|\r)\delta h_j(\hb{\bf x}|\r)\rangle=
B_{tt}(r)\left(\delta_{ij}-n_in_j\right)+B_{rr}n_in_j,
\]
where ${\bf n}=\r/|\r|$. The incompressibility condition implies
that $\ds
B_{tt}=\frac{1}{2r}\partial_r\left(r^2B_{rr}\right)$\cite{landau}.
Then
\begin{equation}
\langle h_i(\hb{\bf x})h_i(\xr)\rangle=\langle \h^2(\hb{\bf
x})\rangle- \frac{1}{2r^2}\partial_r\left(r^3B_{rr}\right)
\label{tens1}
\end{equation}
We represent $g_{ii}$ in the form
\begin{equation}
g_{ii}=\langle a_i(\hb{\bf x})h_i(\xr)\rangle=\langle
a_i(\xr)h_i(\xr)\rangle-\frac{2}{r^2}\partial_r\left(r^3
C(r)\right). \label{tens2}
\end{equation}
Substituting expressions (\ref{tensor})--(\ref{tens2}) into Eqs.
(\ref{eqn1} and (\ref{eqn2}) we obtain
\begin{equation}
-2\bar\varepsilon_m-\frac{1}{2}\partial_t\frac{1}{r^2}\partial_r\left(r^3B_{rr}
\right)= {\ds -\frac{4\hb{f}}{r^2}\partial_r
\left(\frac{1}{r}\partial_r\left(r^5 V\right)\right)
-\frac{\nu_m}{r^2}\partial_r
\left(\frac{1}{r^2}\partial_r\left(r^3B_{rr}\right)\right), }
\end{equation}
\begin{equation}
-2\bar\eta_m-\partial_t\frac{2}{r^2}\partial_r\left(r^3 C\right)
=-\frac{4\hb{f}}{r^2}\partial_r\left(\frac{1}{r}\partial_r\left(r^5
T\right)\right)
-\frac{2\nu_m}{r^2}\partial_r\left(r^2\partial_r\left(
\frac{2}{r^2}\partial_r\left(r^3 C\right)\right)\right).
\end{equation}
Here
\begin{eqnarray}
\bar\varepsilon_m=\nu_m\left\langle\frac{\partial h_i}{\partial
x_j}\frac{\partial h_i}{\partial x_j}\right\rangle=\nu_m\langle
\left(\rot\h\right)^2\rangle,\\
\bar\eta_m= \nu_m\left\langle\frac{\partial a_i}{\partial x_j}
\frac{\partial h_i}{\partial x_j}\right\rangle
=\nu_m\langle\h\rot\h\rangle.
\end{eqnarray}
are, respectively, the dissipation of the magnetic energy and
helicity. Successive integration with allowance for the regularity
\begin{equation}
-\frac{4}{3}\bar\varepsilon_m -\partial_tB_{rr}= {\ds
-\frac{8\hb{f}}{r^4}\partial_r\left(r^5 V\right)
-\frac{2\nu_m}{r^4}\partial_{r} \left(r^4\partial_r B_{rr}\right),
}
\end{equation}
\begin{equation}
{\ds -\frac{\bar\eta_m}{3}-\partial_t C
=-\frac{2\hb{f}}{r^4}\partial_r\left(r^5 T\right)
-\frac{2\nu_m}{r^4}
\partial_r\left(r^4\partial_r C\right).
}
\end{equation}
In the inertial range the time derivatives and dissipation can be
neglected, and it is found that the functions $T$ and $V$ depend
only on the rates of dissipation of the magnetic energy and the
helicity and are, respectively,
\begin{equation}
V=\frac{\bar\varepsilon_m/\hb{f}}{30},\qquad
T=\frac{\bar\eta_m/\hb{f}}{30}
\end{equation}
Therefore the rank-3 two-point correlation tensor for magnetic
field fluctuations becomes
\begin{equation}
\left\langle h_{i}\left(\hb{\bf x}\right) h_{j}\left(\hb{\bf
x}\right) h_{k}\left(\xr\right) \right\rangle
=\frac{\bar\varepsilon_m/\hb{f}}{30} \left( \varepsilon _{jkl}r
_{i} +\varepsilon _{ikl}r _{j}\right)r_l -
\frac{\bar\eta_m/\hb{f}}{10} \left( r _{i}\delta_{jk}+r _{j}\delta
_{ik} - \frac{2}{3}\delta _{ij}r _{k}\right). \label{tensfinal}
\end{equation}

It should be noted especially that up to numerical factors the
tensor (\ref{tensfinal}) of coefficients is identical to the
corresponding correlation tensor of the velocity fluctuations in
hydrodynamic turbulence \cite{lvov97}.

Let us decompose the magnetic field into longitudinal and
transverse components
\begin{eqnarray}
\h_l=\left(\h\r\right)\r/{r^2}, \qquad
\h_t=\h-\h_l,\nonumber\\
\delta h_l\left(\hb{\bf x}|\r\right)=\left(\h_l(\hb{\bf
x+r})-\h_l(\hb{\bf x})\right)\r/r.\nonumber
\end{eqnarray}
In this notation we obtain
\begin{eqnarray}
\langle\delta h_l(\hb{\bf x}|\r)^3\rangle =-24Tr= \langle\delta
\h_l(\hb{\bf x}|\r) [\h_t(\hb{\bf x+ r})\times \h_t({\bf
x})]\rangle=4Vr^2=\frac{2}{15}\bar\varepsilon_m/\hb{f}\cdot
r^2.\nonumber\\
\label{45}
\end{eqnarray}
Therefore the 4/5 and 2/15 laws should hold in homogeneous and
isotropic EMHD turbulence. As one can see from Eq.(\ref{45}), it
is much simpler to determine the helicity in  EMHD turbulence than
in hydrodynamics, where this requires especially accurate
measurements of various velocity components or the use of delicate
instruments to determine  the gradients, whereas in EMHD it is
sufficient to measure only the longitudinal components of the
fluctuational magnetic fields or currents.

We underscore that no dimensional considerations were used to
derive the relations for $T$ and $V$, which involve the helicity
and energy fluxes. This result, which is only a  consequence of
the statistical properties of the isotropic solutions of the EMHD
equations, is universal and does not depend on which kind of
turbulence --- weak or strong ---  develops in the system.

It can be verified by direct substitution that taking the isotropy
into account in the form of an external constant magnetic field
$\h_0=const$ leads only to a modification of the  results
obtained. A dependence on the angle between the radius vector and
the direction  of the magnetic field will appear, since if
homogeneity is preserved, the terms related to  the external field
($\sim (\h_0\nabla)\rot\h$) will not appear in equations of the
form (\ref{eqn1}) and (\ref{eqn2}) for the two-point correlation
functions.

\vskip 5mm {\small
$$
\begin{array}{|c|c|c|}
\hline &
\hb{\bf Hydrodynamics} & \hb{\bf EMHD}\\& \hb{\bf v} & \hb{\bf h} \\
\hline \hb{Nonlinearity} & \ds\left[\rot {\bf v}\times{\bf
v}\right]+\nabla\left(p+\frac{v^2}{2}\right) &
\hb{f}\>\rot\left[\rot \h\times\h\right]
\\
\hline
\begin{array}{c}
\hb{Energy}\\
\hb{dissipation}
\end{array}
& \bar\varepsilon=\nu\langle (\rot\v)^2\rangle &
\bar\varepsilon_m=\nu_m\langle(\rot\h)^2\rangle
\\
\hline
\begin{array}{c}
\hb{Helicity}\\
\hb{dissipation}
\end{array}
& \bar\eta=\nu\langle\rot\v\rot^2\v\rangle &
\bar\eta_m=\nu_m\langle\h\rot\h\rangle
\\
\hline
\hb{Kolmogorov scaling} &
E(k)=C\bar\varepsilon^{2/3}k^{-5/3} &
E_m(k)=C_m\left(\bar{\varepsilon}_m/\hb{f}\right)^{2/3}k^{-7/3}
\\
\hline \hb{Helical scaling} & E(k)=C^*\bar\eta^{2/3}k^{-7/3} &
E_m(k)=C^*_m\left(\bar\eta_m/\hb{f}\right)^{2/3}k^{-5/3}
\\
\hline \hb{$4/5$ law} & \langle\delta v_l(\hb{\bf x}|\r)^3\rangle
=-{\ds{4\over 5}}\bar\varepsilon\cdot r & \ds\langle\delta
h_l(\hb{\bf x}|\r)^3\rangle =-{\ds{4\over
5}}\bar\eta_m/\hb{f}\cdot r
\\
\hline \hb{$2/15$ law} &
\begin{array}{c}
\langle\delta\v_l(\hb{\bf x}|\r)\left[\v_t(\xr)\times \v_t(\hb{\bf
x})\right]\rangle=
\\
{\ds\frac{2}{15}}\bar\eta\cdot r^2
\end{array}
&
\begin{array}{c}
\langle\delta \h_l(\hb{\bf x}|\r)[\h_t(\xr)\times
\h_t(\hb{\bf x})]\rangle=\\
{\ds\frac{2}{15}}\bar\varepsilon_m/\hb{f}\cdot r^2
\end{array}
\\
\hline
\end{array}
$$
}

The "extra" curl in EMHD, as compared with the Navier-Stokes
equation, leads to an unusual transposition --- for longitudinal
correlations the 4/5 law holds, just as in  hydrodynamics, but it
is related with the gyrotropic component of the fluctuations,
i.e.,  the helicity flux and, conversely, mixed
longitudinal-transverse correlations are related  with the
magnetic energy flux. Table I gives a comparative summary of the
basic results.

In closing, I thank S. S. Moiseev for helpful discussions. This
work was supported in part by the Russian Fund for Fundamental
Research (Grant No. 98-02-17229) and  INTAS (Joint Georgia-INTAS
Project No. GE-504).

*e-mail: ochkheti@mx.iki.rssi.ru


\begin{thebibliography}{99}

\bibitem{kolm41} A. N. Kolmogorov. Dokl. Akad. Nauk SSSR \textbf{32}, 19 (1941).

\bibitem{chandras} Chandrasekhar. Proc. Phys. Soc. London. Sect. A \textbf{204}. 435
(1951).

\bibitem{chkhet96} O. G. Chkhetiani. JETP Lett. \textbf{63}. 808 (1996).

\bibitem{lvov97} V. S. L'vov, E. Podivilov. and I. Procaccia.
http://xxx.lanl.gov/abs/chao-dyn/9705016.

\bibitem{Monin96} A. S. Monin and A. M. Yaglom. \emph{Statitstical Fluid Mechanics}. Vols.
1 and 2 (MIT Press, Cambridge. Mass.. 1971 and 1975) [Russian
original, Gidrometeoizdat, St. Petersburg. 1996, Part 2].

\bibitem{biferale} L. Biferale. D. Pierotti. and P. Toschi.
http://xxx.lanl.gov/abs/chao-dyn/9804004.

\bibitem{frisch95}  U. Frisch, \emph{Turbulence}, Cambridge University Press, New
York. 1995.

\bibitem {kingsep87} A. S. Kingscp. K. V. Chukbar, and V. V. Yan'kov.
\emph{Reviews of Plasma Physics}, Vol. \textbf{16}, edited by
B.B.Kadomtsev, Consultants Bureau. New York (1990) [Russian
original. Voprosy Teorii Plazmy \textbf{16}. 209 (1987)].

\bibitem {gordeev94} A. V. Gordeev, A. S. Kingsep. and L. I. Rudakov, Phys. Rep. \textbf{243}.
216 (1994). '\r{ }

\bibitem {vainshtein76} S. I. Vainshteln. Usp. Fiz. Nauk \textbf{120}. 613 (1976) [Sov. Phys. Usp.
19. 987 (1976)].

\bibitem {yakovenko69} V. M. Yakovenko, Zh. Eksp. Teor. Fiz. \textbf{57}. 554 (1968) [sic].

\bibitem {cytovich} V. N. Tsytovich, \emph{Theory of Turbulent Plasma} (Plenum Press,
New York, 1974) [Russian original. Energoatomizdat, Moscow. 1971].

\bibitem {livshitz72} M. A. Livshits and V. N. Tsytovich, Zh. Eksp. Teor. Fiz. \textbf{6}2. 606
(1972) [Sov. Phys. JETP \textbf{35}, 321 (1972)].

\bibitem {vainshtein73} S. I. Vainshtein. Zh. Eksp. Teor. Fiz. \textbf{64}. 139 (1973) [Sov. Phys.
JETP \textbf{37}. 73 (1973)].

\bibitem{landau} L. D. Landau and E. M. Lifshitz. F/``M AfecAamci, 2nd edition
(Pergamon Press, New York. 1987) [Russian original. 3rd ed.,
Nauka, Moscow, 1986].

\end{thebibliography}
\end{document}